\documentclass[twocolumn,showpacs,preprintnumbers,amsmath,amssymb]{revtex4}
\usepackage{dcolumn}
\usepackage{bm}
\usepackage{graphicx}

\newcommand{\ket}[1]{\displaystyle{|#1\rangle}}
\newcommand{\bra}[1]{\displaystyle{\langle #1|}}

\newcommand{\sumkj}{\sum_{\mathbf{k}j}}

\newcommand{\sumkkjj}{\sum_{\mathbf{k}\mathbf{k'}jj'}}
\newcommand{\akj}{a_{\mathbf{k}j}}
\newcommand{\ackj}{a^{\dag}_{\mathbf{k}j}}
\newcommand{\akjp}{a_{\mathbf{k}'j'}}

\newcommand{\omegak}{\omega_k}
\newcommand{\omegakp}{\omega_{k'}}
\newcommand{\omegaz}{\omega_{0}}
\newcommand{\fkjR}{\mathbf{f}(\mathbf{k}j,\mathbf{r})}
\newcommand{\fkjpR}{\mathbf{f}(\mathbf{k'}j',\mathbf{r})}

\newcommand{\Sp}{S_{+}}
\newcommand{\Sm}{S_{-}}
\newcommand{\Sz}{S_{z}}

\newcommand{\ekjx}{(\hat{e}_{\mathbf{k}j})_{x}}
\newcommand{\ekjy}{(\hat{e}_{\mathbf{k}j})_{y}}
\newcommand{\ekjz}{(\hat{e}_{\mathbf{k}j})_{z}}

\newcommand{\fkjx}{(\mathbf{f}_{\mathbf{k}j})_{x}}
\newcommand{\fkjy}{(\mathbf{f}_{\mathbf{k}j})_{y}}
\newcommand{\fkjz}{(\mathbf{f}_{\mathbf{k}j})_{z}}

\begin{document}
\preprint{APS/123-QED}
\title{Dynamical Casimir-Polder force between an atom and a conducting wall}

\author{R. Vasile\mbox{${\ }^{1}$} and R. Passante\mbox{${\ }^{2}$}}
\affiliation{ \mbox{${\ }^{1}$} Department of Physics and
Astronomy, University
of Turku, FIN-20014 Turun Yliopisto, Finland \\
\mbox{${\ }^{2}$} Dipartimento di Scienze Fisiche ed Astronomiche
dell'Universit\`{a} degli Studi di Palermo and CNSIM,\\Via
Archirafi 36, I-90123 Palermo, Italy}

\begin{abstract}
The time-dependent Casimir-Polder force arising during the time
evolution of an initially bare two-level atom, interacting with
the radiation field and placed near a perfectly conducting wall,
is considered. Initially the electromagnetic field is supposed to
be in the vacuum state and the atom in its ground state. The
analytical expression of the force as a function of time and
atom-wall distance, is evaluated from the the time-dependent
atom-field interaction energy. Physical features and limits of
validity of the results are discussed in detail.
\end{abstract}


\maketitle

\section{Introduction}

The presence of vacuum fluctuations is a remarkable property of
quantum field theory. In quantum electrodynamics these
fluctuations are responsible of many observable effects such as
Casimir and Casimir-Polder forces
\cite{Casimir48,CP48,Milonni94,CPP95}. They are long-range
electromagnetic interactions between neutral objects such as
atoms/molecules or macroscopic bodies. These interactions are
related to the properties of the vacuum state of the
electromagnetic field, in particular to the fact that vacuum field
fluctuations may polarize the atoms or exert a pressure on
macroscopic bodies, and thus induce electromagnetic interactions.
Recently, Casimir forces have been measured with remarkable
precision for different topologies \cite{Lamoreaux05,Onofrio06}.
The atom-wall Casimir-Polder force has also been measured with
precision \cite{SBCS93,DD03}, and a good agreement with
theoretical predictions has been obtained.

In this paper we consider the Casimir-Polder force between an atom
and a perfectly conducting plate in a dynamical (i.e.
time-dependent) situation. Recent papers have given special
attention on the problem of time-dependent Casimir-Polder forces
between two or more atoms or dynamical Casimir forces between
macroscopic objects \cite{RPP04,RPP07,BKWD04,BMM01}. Current
experiments are trying to detect the real photons emitted in the
dynamical Casimir effect \cite{ABBC08}. In this paper we consider
the dynamical Casimir-Polder force arising between a perfectly
conducting plate and an initially bare ground-state atom during
its dynamical self-dressing. We obtain an analytical expression of
the time-dependent Casimir-Polder force, which is characterized by
a timescale corresponding to the time taken by the field emitted by
the atom to go back to the atomic position, after reflection
on the conducting plate. This is physically sound, because after
this time the interaction of the atom with the (reflected)
reaction field starts. We also find that the dynamical
Casimir-Polder force can be attractive or repulsive according to
time and atom-wall distance (on the contrary, the stationary
Casimir-Polder force for a ground-state atom is attractive for any atom-wall separation
\cite{CP48}).

This paper is organized as follows. In Section \ref{sec:2} we
introduce the Hamiltonian describing our system in the multipolar
coupling scheme and solve the relevant Heisenberg equations for atomic and field operators. In Section \ref{sec:3} we discuss a method for
obtaining the stationary atom-wall Casimir-Polder energy and
force, which is then generalized to the time-dependent case that
is the main subject of the paper. In Section \ref{sec:3} we also
discuss the results obtained and their physical interpretation, as
well as possible future developments.

\section{The Hamiltonian and the Heisenberg equations}
\label{sec:2}

We consider a two-level atom in front of an infinitely extended
and perfectly conducting wall placed at $z=0$. In the multipolar
coupling scheme and within dipole approximation, our system is
described by the following Hamiltonian \cite{CPP95}

\begin{equation}\label{eq:2.1}\begin{split}
H&=H_0+H_{I}\\
H_0&=\hbar\omegaz\Sz+\sumkj\hbar\omegak\ackj\akj\\
\end{split}\end{equation}
\begin{equation}\label{eq:2.2}\begin{split}
H_I&=- i\sqrt{\frac{2\pi\hbar c}{V}}\sumkj\sqrt{k}
\bigl[\mbox{\boldmath $\mu$}\cdot\fkjR\bigl]\\
&\times(\Sp\akj+\Sm\akj-\Sp\ackj-\Sm\ackj)
\end{split}\end{equation}
where $\omegaz=ck_0$ is the transition frequency between the
atomic levels and $\fkjR$ are the field mode functions evaluated
at the atomic position $\mathbf{r}$, that take into account the
presence of the wall. $\Sz,\Sp,\Sm$ are the pseudo-spin operators
of the two-level atom, and $\mbox{\boldmath $\mu$}$ is its electric dipole
moment operator.

In the presence of an infinite perfectly conducting wall placed at $z=0$,
the mode functions $\fkjR$ are \cite{Milonni94,PT82}
\begin{widetext}
\begin{equation}\label{eq:2.3}\begin{split}
&\fkjx=\sqrt{8}\ekjx\cos\biggl[k_{x}\biggl(x+\frac L2\biggl)\biggl]
\sin\biggl[k_{y}\biggl(y+\frac L2\biggl)\biggl]\sin\bigl(k_{z}z\bigl)\\
&\fkjy=\sqrt{8}\ekjy\sin\biggl[k_{x}\biggl(x+\frac L2\biggl)\biggl]
\cos\biggl[k_{y}\biggl(y+\frac L2\biggl)\biggl]\sin\bigl(k_{z}z\bigl)\\
&\fkjz=\sqrt{8}\ekjz\sin\biggl[k_{x}\biggl(x+\frac L2\biggl)\biggl]
\sin\biggl[k_{y}\biggl(y+\frac L2\biggl)\biggl]\cos\bigl(k_{z}z\bigl)
\end{split}\end{equation}
\end{widetext}
These expressions of the field modes should be considered in the
limit $L\rightarrow\infty$.

In order to evaluate in the next Section the dynamical
Casimir-Polder energy, we need expressions of field and atomic
operators in the Heisenberg representation. Thus we can write the
Heisenberg equations for atomic and field operators, and solve
them iteratively at the lowest significant order. After
straightforward algebra, we obtain the following expressions for
the Heisenberg operators relevant for the calculations in Section
\ref{sec:3}
\begin{widetext}
\begin{equation}\label{eq:2.4}\begin{split}
&\akj^{(0)}=e^{-i\omegak t}\akj(0)
\qquad\qquad\Sp^{(0)}=e^{i\omegaz t}\Sp(0) \\
&\akj^{(1)}(t)=e^{-i\omegak t}\sqrt{\frac{2\pi\omegak}{\hbar V}}
(\mbox{\boldmath $\mu$}\cdot\fkjR)\left[\Sp(0) F(\omegaz+\omegak,t)+\Sm(0)
F(\omegak-\omegaz,t)\right]\\
&\Sp^{(1)}(t)=-2\Sz(0) e^{i\omegaz
t}\sumkj\sqrt{\frac{2\pi\omegak}{\hbar
V}}(\mbox{\boldmath $\mu$}\cdot\fkjR)\left[\akj(0)
F^*(\omegaz+\omegak,t)-\ackj(0) F(\omegak-\omegaz,t)\right]
\end{split}\end{equation}
\end{widetext}
(the superscripts indicate the perturbative order), where we have introduced the auxiliary function
\begin{equation}\label{eq:2.5}
F(x,t)=\int_{0}^{t}e^{ixt'}dt'=\frac{e^{ixt}-1}{ix}
\end{equation}

\section{The dynamical Casimir-Polder interaction energy}
\label{sec:3}

In order to obtain the dynamical Casimir-Polder force, let us
first consider the stationary case. In this case, the atom-wall
potential energy for a ground state atom is obtained from the
second-order energy shift $\Delta E^{(2)}$ of the bare ground
state $\ket{0,\downarrow}$ (field in the vacuum state and atom in
its ground state) due to the atom-field interaction, calculated
by perturbation theory. The force is then obtained as minus the
derivative of the potential energy with respect to the atom-wall
distance $d$
\begin{equation}\label{eq:3.1}
F(d)=-\frac{\partial (\Delta E^{(2)})}{\partial d}
\end{equation}
In this quasi-static approach, the atom's translational degrees of
freedom are not taken into account. The well-known result for an
isotropic two-level atom is \cite{Barton87}
\begin{equation}\label{eq:3.2}\begin{split}
F_{g}(d,k_0)&=-\frac{\mu^2}{12\pi
d^4}\Big( 8k_0d-6(2k_0^2d^2-1)f(2k_0d)\\
&-4k_0d(2k_0^2d^2-3)g(2k_0d)\Big)
\end{split}\end{equation}
where $f(z)$ and $g(z)$ are the auxiliary functions of the sine
and cosine integral functions \cite{AS64}. This force is negative
for any atom-wall distance $d$, yielding an attractive force, and
behaves as $d^{-4}$ for $d \ll k_0^{-1}$ (near zone) and as
$d^{-5}$ for $d \gg k_0^{-1}$ (far zone) \cite{CP48}.

It is easy to show, independently from the explicit form of
$H_I$, that
\begin{equation}\label{eq:3.3}
\Delta E^{(2)}= \frac{ {\ }_D
\bra{0,\downarrow}H_I\ket{0,\downarrow}_D}2
\end{equation}
where $\ket{0,\downarrow}_D$ is the first-order {\it dressed}
ground state of the system
\begin{equation}\label{eq:3.4}
\ket{0,\downarrow}_{D}=\ket{0,\downarrow}-\sum_{\ket{\psi}\neq\ket{0,\downarrow}}
\frac{\bra{\psi}H_{I}\ket{0,\downarrow}}{E_{\psi}-E_0}\ket{\psi}
\end{equation}
This relation shows that the second-order energy shift can be also
obtained from the average value of the interaction Hamiltonian on
the dressed ground state of the system \cite{MPRSV08}.

We now consider the time-dependent situation, when the atom is
initially in its bare ground-state. This state is not an
eigenstate of the total Hamiltonian and thus it evolves in time
(dynamical self-dressing) \cite{CPP88,PPP93}, finally yielding a
time-dependent Casimir-Polder force between the atom and the wall.
In order to obtain the dynamical (i.e. time-dependent)
Casimir-Polder energy-shift, and then the dynamical force, we can
use a generalization of \eqref{eq:3.3}. In fact, we shall evaluate
$H_I(t)/2$ in the Heisenberg representation and its average value
on the initial state, that is the bare ground state of the
atom-wall system. Using the expressions \eqref{eq:2.4} for the
Heisenberg operators into the expression \eqref{eq:2.2} of the
interaction Hamiltonian, at the second order in the atom-field
interaction we obtain
\begin{widetext}
\begin{equation}\begin{split}\label{eq:3.5}
&H_I^{(2)}(t)=-\frac{2\pi ic}V\sumkj k\big|\mbox{\boldmath $\mu$}\cdot\fkjR\big|^2
\bigg(\Sp(0) e^{i\omegaz t}+h.c.\bigg)  \biggl[
\Sp(0)\bigg(e^{-i\omegak t}F(\omegaz+\omegak,t)-e^{i\omegak t}F^*(\omegak-\omegaz,t)\biggl)-h.c.\biggl]\\
&+\frac{4\pi
ic}V\Sz(0)\sumkkjj\sqrt{kk'}\left(\mbox{\boldmath $\mu$}\cdot\fkjR\right)\left(\mbox{\boldmath $\mu$}\cdot\fkjpR\right)\biggl[
\akjp(0)\biggl(e^{i\omegaz t}F^*(\omegaz+\omegakp,t)-e^{-i\omegaz t}F^*(\omegakp-\omegaz,t)\biggl)+h.c.\biggl]\\
&\times \biggl(\akj(0) e^{-i\omegak t}-h.c.\biggl)
\end{split}\end{equation}
\end{widetext}

In analogy with the stationary case outlined above, we can
evaluate the time-dependent Casimir-Polder atom-wall energy shift as
\begin{widetext}
\begin{equation}\begin{split}\label{eq:3.6}
\Delta E^{(2)}(d,t)&=
\frac{\bra{0,\downarrow}H_{I}^{(2)}(t)\ket{0,\downarrow}}2 \\
&=-\frac{i\pi c}{V}\sumkj \left(\mbox{\boldmath $\mu$}\cdot\fkjR\right)^2 \Big(
F^*(\omegaz+\omegak,t)e^{i(\omegaz+\omegak)t}-F(\omegaz+\omegak,t)e^{-i(\omegaz+\omegak)t}\Big)
\end{split}\end{equation}
\end{widetext}
A similar approach for obtaining a time-dependent Casimir-Polder
energy has already been used to calculate the dynamical
Casimir-Polder force between a ground-state and an initially bare
excited atom \cite{RPP04} as well as dynamical three-body
Casimir-Polder forces \cite{RPP07}. The expression \eqref{eq:3.6}
can be explicitly evaluated at any time $t>0$ compatible with our
second-order perturbative expansion, and for any atom-wall
distance $d$.

In the continuum limit, after some algebraic manipulation, eq. \eqref{eq:3.6} becomes
\begin{equation}\label{eq:3.7}\begin{split}
\Delta E^{(2)}(d,t)&=-\frac{\mu^2}{4\pi
d^3}\int_{0}^{\infty}\frac{-2x\cos(x)+(2-x^2)\sin(x)}{x+x_0}\\
\times&\bigl(1-\cos[a(x+x_0)]\bigl)dx
\end{split}\end{equation}
where $x_0=2k_0d$ and $a=\frac{ct}{2d}$. This expression can be
written in the more compact form
\begin{equation}\label{eq:3.8}
\Delta E^{(2)}(d,t)=\lim_{m\rightarrow 1}\biggl[ D_{m}
\int_{0}^{\infty}\frac{\sin(mx)}{x+x_0}\bigl(1-\cos[a(x+x_0)]\bigl)dx
\biggl]
\end{equation}
where we have defined the differential operator $D_{m}$ as
\begin{equation}\label{eq:3.9}
D_{m}=-\frac{\mu^2}{4\pi d^3}\biggl[2-2\frac{\partial}{\partial
m}+\frac{\partial^2}{\partial m^2}\biggl]
\end{equation}

A few words are necessary for the $x-$integral in \eqref{eq:3.7}.
The energy shift $\Delta E^{(2)}(d,t)$ diverges for $a=1$, that is for
$t=2d/c$, which is the time at which the field emitted by the
atom, after reflection on the conducting wall, goes back at the
atomic position. This divergence is not surprising, being related
to the well-known divergences of the radiation reaction field and
to the dipole approximation \cite{Milonni76}. It is also related
to the assumption of an initially bare state \cite{PV96,PP03}. For
this reason, we shall not consider the interaction energy at times
$t\sim 2d/c$. It is however worth to consider the Casimir-Polder
energy for $t < 2d/c$ and for $t > 2d/c$ ($a<1$ and $a>1$,
respectively), that is before and after the ``back-reaction'' time.
In both these timescales, the analytical expression for the
Casimir-Polder force, obtained in a quasi-static approach as
minus the spatial derivative of the interaction energy, is
\begin{widetext}
\begin{equation}\begin{split}\label{eq:3.10}
&F(d,t,k_0)= -\frac{\partial(\Delta E^{(2)})(d,t)}{\partial d}=
-\frac{\mu^2}{12\pi
d^4}\biggl[8k_0d+\frac{16d^3k_0(c^2t^2-8d^2)\cos(ck_0t)}{(c^2t^2-4d^2)^2}+\\
&-\frac{4dct\sin(k_0ct)}{(4d^2-c^2t^2)^3}\bigl[16d^4
\bigl(2k_0^2d^2-9\bigl)-16c^2t^2d^2\bigl(k_0^2d^2-2\bigl)+c^4t^4\bigl(2k_0^2d^2-3\bigl)\bigl]+\\
&+\bigl(Ci[k_0(2d+ct)]-2Ci(2k_0d)+Ci[k_0|2d-ct|]\bigl)\bigl[2k_0d
\cos(2k_0d)\bigl(3-2k_0^2d^2\bigl)-3\bigl(1-2k_0^2d^2\bigl)\sin(2k_0d)\bigl]+\\
&+\bigl(Si[k_0(2d+ct)]-2Si(2k_0d)+Si[k_0(2d-ct)]+l\pi\bigl)\bigl[3\bigl(1-2k_0^2d^2\bigl)
\cos(2k_0d)+2k_0d\bigl(3-2k_0^2d^2\bigl)
\sin(2k_0d)\bigl]\biggl]
\end{split}\end{equation}
\end{widetext}
where $Si(x)$ and $Ci(x)$ are respectively the sine- and cosine
integral functions \cite{AS64}. The parameter $l$ in \eqref{eq:3.10} is $0$ for $a<1$
and $1$ for $a>1$. This expression clearly shows the divergence at
$t=2d/c$ discussed above.

Let now discuss some features of physical relevance of the
expression obtained for the dynamical Casimir-Polder force. At
$t=0$ the force vanishes, due to the initial condition of a bare
state. For successive times the force increases with an
oscillatory behavior both in time and space, with scales given by
$k_0$. Depending on the time $t$ and the atom-wall distance $d$,
the force can be positive (attractive) or negative (repulsive).
This is a new feature compared to the stationary case for the
ground-state atom, where the force is always attractive. Figure
\ref{Fig:1} shows a plot of the Casimir-Polder force for a fixed
value of $d$ as a function of time for $a<1$, that is $ct<2d$. The
mentioned oscillations of the value of the force are evident. An
oscillatory behaviour of the force occurs also for $a>1$, that is
for $ct>2d$, as shown in Figure \ref{Fig:2}. This plot shows also
that for large $t$ the force settles to a negative value,
asymptotically yielding a static attractive force. From
\eqref{eq:3.8} or \eqref{eq:3.10} it is easy to show that the
asymptotic value of the force coincides with that obtained with a
time-independent approach. This indicates that, at large times,
the atom is fully dressed and thus its interaction with the wall
is the same as in the static case. It is also worth mentioning
that the fact that the force is not zero for $t<d/c$, i.e. before
a light signal from the atom can reach the wall, should not
surprise. In fact, for $t>0$ the atom immediately ``knows'' of the
existence of the wall, because it interacts with the field modes
\eqref{eq:2.3} which incorporate the presence of the conducting
wall.

\section{Conclusion}

In this paper we have considered the dynamical (time-dependent)
Casimir-Polder force between an initially bare ground-state atom
and a perfectly conducting wall. An analytical expression of the
time-dependent force has been obtained, which shows that,
contrarily to the well-known stationary case of an attractive
force for any atom-wall distance, the force oscillates in time
from attractive to repulsive values. The scale of these
oscillations is related to the atomic transition frequency. A
characteristic timescale of the dynamical Casimir-Polder force is
shown to be twice the time taken by a light signal to cover the
atom-wall distance. This is physically understandable, because
this is the time taken by the field emitted by the atom to go back
to the atomic position, after being reflected on the wall, and
interact with its source. Work in progress concerns the dynamical
atom-wall Casimir-Polder force in the case of an initially excited
atom.

\begin{acknowledgments}
Partial support by Ministero dell'Universit\`{a} e della Ricerca
Scientifica e Tecnologica and by Comitato Regionale di Ricerche
Nucleari e di Struttura della Materia is also acknowledged.
\end{acknowledgments}

\begin{widetext}

\newpage

\begin{figure}[ht]
\begin{center}
\includegraphics*[width=16cm]{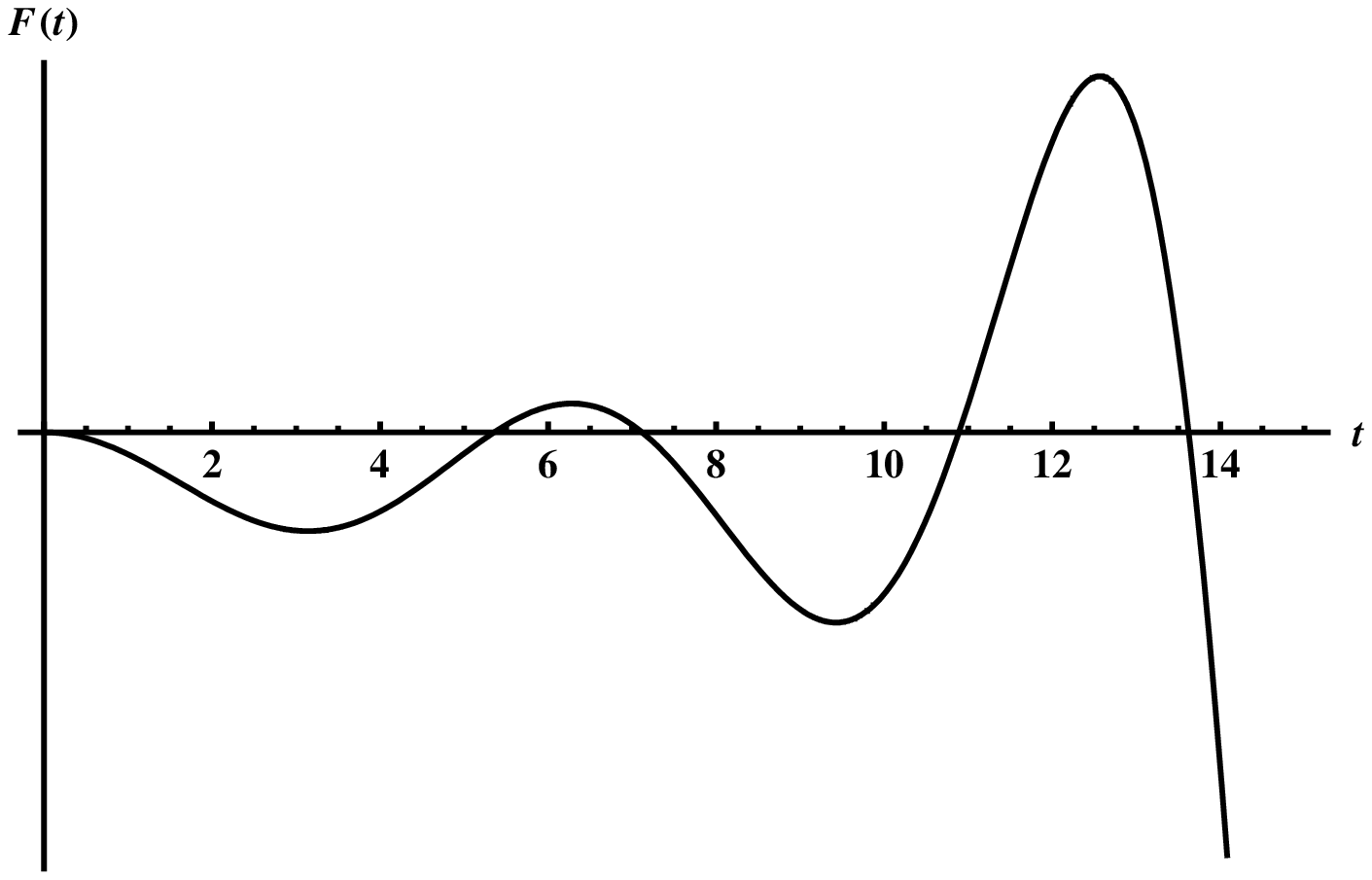}
\end{center}
\caption{The dynamical Casimir-Polder force for $t<2d/c$, that is
before the back-reaction time. Units are such that $c=1$ and
$k_0=1$. The atom-wall distance is $d=10$, thus the back-reaction time is $t=20$.} \label{Fig:1}
\end{figure}

\newpage

\begin{figure}[ht]
\begin{center}
\includegraphics*[width=16cm]{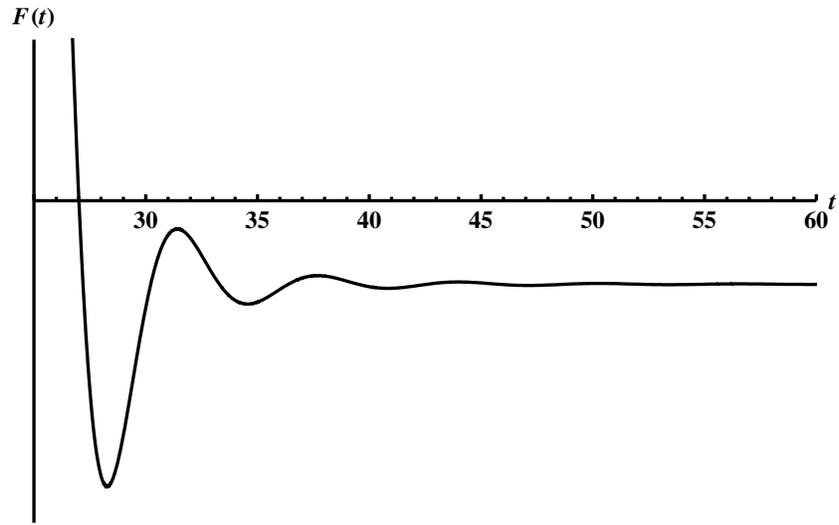}
\end{center}
\caption{The dynamical Casimir-Polder force for $t>2d/c$, that is
after the back-reaction time. Same units and parameters of Fig. 1 have been used.} \label{Fig:2}
\end{figure}

\end{widetext}

\end{document}